\documentclass{article}
\usepackage[T1]{fontenc}
\usepackage[latin9]{inputenc}
\usepackage[letterpaper]{geometry}
\usepackage{textcomp}
\usepackage{amstext}
\usepackage{graphicx}
\usepackage{esint}
\usepackage[authoryear]{natbib}

\makeatletter
\newcommand{\lyxaddress}[1]{
	\par {\raggedright #1
	\vspace{1.4em}
	\noindent\par}
}

\makeatother

\begin{document}
\title{The range of Jupiter's flow structures that fit the Juno asymmetric gravity measurements}
\author{Keren Duer$^{1}$, Eli Galanti$^{1}$
 and Yohai Kaspi$^{1}$ \\
\\
(JGR planets, in rev.)}
\maketitle

\lyxaddress{\begin{center}
\textit{$^{1}$Department of Earth and Planetary Sciences, Weizmann
Institute of Science, Rehovot, Israel. }\\
\par\end{center}}

\begin{abstract}
The asymmetric gravity field measured by the Juno spacecraft has allowed the
estimation of the depth of Jupiter's zonal jets, showing
that the winds extend approximately $3000$~km beneath the cloud level.
This estimate was based on an analysis using
a combination of all measured odd gravity harmonics, $J_{3}$, $J_{5}$,
$J_{7}$, and $J_{9}$, but the wind profile's dependence on each
of them separately has yet to be investigated. Furthermore, these calculations
assumed the meridional profile of the cloud-level wind extends to
depth. However, it is possible that the interior jet profile varies somewhat
from that of the cloud level. Here we analyze in detail the possible
meridional and vertical structure of Jupiter's deep
jet-streams {that can match the gravity measurements}. We find that each odd gravity harmonic constrains the
flow at a different depth, with $J_{3}$ the most dominant at
depths below $3000$~km, $J_{5}$ the most restrictive overall,
whereas $J_{9}$ does not add any constraint on the flow if the other odd harmonics are considered. Interior flow profiles constructed
from perturbations to the cloud-level winds allow a more extensive
range of vertical wind profiles, yet when the meridional profiles differ substantially
from the cloud level, the ability to match the gravity data
significantly diminishes. Overall, we find that while interior wind profiles
that do not resemble the cloud level are possible, they are statistically
unlikely. {Finally, inspired by the Juno microwave radiometer measurements, assuming the brightness temperature is dominated by the ammonia abundance, we find that depth-dependent flow profiles are still compatible with the gravity measurements.}
\end{abstract}

\section*{Plain Language Summary}

Jupiter's north-south asymmetric gravity field, as measured by
the Juno spacecraft, currently orbiting Jupiter, has been used to set the 
depth of its jet-streams (associated with the famous visible cloud
bands) at approximately $\sim3000$ km. This estimate was based on
all the gravity field measurements combined. However, there is also
information about the structure of the flow hidden in each individual measurement. 
Here we analyze these measurements and show how each of them
constrains the flow at a different depth. We also systematically investigate
the statistical likelihood of wind profiles that differ from the profile
observed at the cloud level with various structures at depth. {We find that Jupiter's measured cloud-level jet streams fit with its gravity data only for a relatively narrow envelope of vertical structures.}
Although other jet profiles that are different from the one observed at
the cloud level are feasible (still consistent with the gravity data),
they are statistically unlikely. Finally, we explore a depth-dependent
wind structure inspired by the Juno microwave radiometer instrument,
which indicates that ammonia abundance varies with depth and might
be correlated with the jet-streams. We find that such a profile can
still match the gravity data as long as the variation from the cloud-level
wind is not substantial.

\section{Introduction\label{sec:Introduction}}

The Juno spacecraft has provided an unprecedented glance into Jupiter's
atmospheric flows below the cloud level. The high-precision gravity
measurements, particularly those of the odd gravitational harmonics repeated
in multiple passes \citep{Iess2018}, have presented an opportunity to
estimate the depth and structure of Jupiter's zonal jets. It has been found
that the zonal jets are deep and penetrate to approximately $3000$~km
below the cloud level \citep{Kaspi2018}. Below this depth, the even
gravitational harmonics indicate that Jupiter rotates almost like
a solid body \citep{Guillot2018}. However, determining the details
of the decay profile with depth poses a significant challenge. Remnants of the zonal flows appear even below $4000$~km, {and
since the estimation of the electrical conductivity in Jupiter at
this depth is at least} {$10\;{\rm S~m^{-1}}$}
\citep{nellis1996,wicht2019a,wicht2019b}, an interaction between the
flow and the magnetic field is expected there \citep{Cao2017,Galanti2017c,duer2019,moore2019}.
Understanding the gravity harmonic signature and the flow structure
below the cloud level is {thus essential} {in order to} build a better picture {of Jupiter's atmosphere}.

{The gravity field of Jupiter, {represented by the gravity harmonics}, reflects both the internal density
structure and the deep zonal flow structure \citep{Hubbard1999,Kaspi2010a}.}
The even gravity harmonics are used to constrain the internal density
structures of Jupiter and other gas giants \citep[e.g.,][]{hubbard1974,Hubbard1975a,helled2010c,Nettelmann2013}.
Multiple studies have shown that the higher-order (even) gravity harmonics
are sensitive to the outer regions of the planet \citep[e.g.,][]{zharkov1974,Guillot2007,Nettelmann2013b}.
Their exact value is defined by the density distribution throughout
the planet and the planet's rotation, composition, shape, mass, and
radius. Since for a static gas planet, the odd harmonics are identically
zero, any gravitational asymmetry between north and south would indicate a dynamical source generating those asymmetries \citep{Kaspi2013a}.
Juno measured with high precision the gravity harmonics up to $J_{10}$,
including significant odd values. The measured values and error range
are: $J_{3}=\left(-4.24\pm0.91\right)\times10^{-8}$, $J_{5}=\left(-6.89\pm0.81\right)\times10^{-8}$,
$J_{7}=\left(12.39\pm1.68\right)\times10^{-8}$ and $J_{9}=\left(-10.58\pm4.35\right)\times10^{-8}$
\citep{Iess2018}. The relation between the density anomaly and the
flow (thermal wind balance) allows constraining the deep flow structure
within the planet \citep{Kaspi2010a,Kaspi2013a,Kaspi2018}. Assuming
that the cloud-level zonal wind profile is extended towards Jupiter's
interior using a scaling factor, one can find many solutions for the
deep flow structure that satisfy all four odd gravity harmonics within
the uncertainty range. {With the currently available data,
Jupiter's deep flow cannot be determined uniquely \citep{Kaspi2018,Kong2018},
and systematic exploration of the range of the deep flow structure
is necessary.}

{Moreover}, the meridional profile of the zonal wind is not
necessarily constant with depth. The cloud-level wind itself has a
measurement error \citep{garcia2001,Salyk2006,Tollefson2017}, and
as it extends inward, the profile might vary, although any such variation
must be accompanied with a meridional temperature gradient as well.
Some evidence for such meridional variations come from the Juno microwave
radiometer (MWR) measurements showing that the nadir brightness temperature
profile (dominated by the ammonia abundance) becomes smoother with
depth \citep{Bolton2017,Li2017}. {Although this measurement does not
necessarily correlate with temperature, it does coincide, to some
degree, with the zonal wind profile at the cloud level \citep{Bolton2017},
and thus might hint to the vertical variation of the zonal
wind profile in the upper 300~km.}

Previous work on constraining the deep flow structure was done using
all four measured gravity harmonics combined \citep[e.g.,][]{Kaspi2018,Kong2018}.
However, an important question is how does each gravity harmonic individually
constrain the flow strength at different depths. Here, we examine
the individual contribution of each odd gravity harmonic, with emphasis
on the depth of influence and the relation to the cloud-level zonal
wind profile. In order to provide a systematic analysis, we take a
hierarchal approach, in which we increase the level of complexity of the
variation of the wind structure, and in all cases explore what is
the range of solutions that match the gravity measurements. We begin
with solutions that are identical to the cloud-level profile and
allow only for the vertical decay to vary. Then, we relax the constraint
on the meridional profile of the zonal wind and allow variations
from the measured cloud-level profile along with the varying vertical
decay. Finally, we examine random meridional profiles that are not
related at all to Jupiter's measured cloud-level profile and explore
the possibility that the interior wind structure, which influences
the gravity measurements, is completely different from the cloud-level
flow. Following this logic, we also search for solutions with smoother
wind profiles that resemble the MWR measurements at 300~km
(channel 1), and calculate the vertical profile of such flows that
can match also the gravity data.

The paper is organized as follows: In section \ref{sec:Methodology},
we introduce the theoretical background for this analysis, connecting
the gravity measurements and the wind profile. In section \ref{sec:vertical_structure},
we present the possible solutions for Jupiter's wind profile, the
depth sensitivity obtained by excluding a specific harmonic, and the
contribution function of each harmonic. In section \ref{sec:meridional},
we discuss the ability to find solutions for the anomalous gravity
field of different meridional profiles, and in section \ref{sec:MWR},
we explore depth-dependent meridional structures, inspired by the
MWR measurements. We discuss the dynamical implications of the results and conclude in section \ref{sec:Conclusions}.

\section{Methodology \label{sec:Methodology}}

The density distribution within Jupiter is reflected in the zonal
gravity harmonics $\left(J_{n}\right)$, which describe the external
gravitational field of the planet in equilibrium \citep{zharkov1974}.
The gravity harmonics can be represented by
\begin{eqnarray}
J_{n} & = & -\frac{1}{MR_{J}^{n}}\int\rho r^{n}P_{n}\left(\mu\right)d^{3}r,\label{eq:Jn}
\end{eqnarray}
where $M$ and $R_{J}$ are Jupiter's mass and equatorial radius,
respectively, $n$ is the harmonic degree {($n=2,...,N$)},
$\rho$ is the density, $r$ is the radial coordinate and {$P_{n}\left(\mu\right)$ is
the $n$-th Legendre polynomial, where $\mu=\sin\theta$ and $\theta$
is the latitude} \citep{Hubbard1984}. The density
can be decomposed such that $\rho\left(r,\theta\right)=\widetilde{\rho}\left(r,\theta\right)+\rho'\left(r,\theta\right)$,
where $\widetilde{\rho}\left(r,\theta\right)$ is the static component
that is determined by the planet's shape and rotation \citep{Hubbard2012},
and $\rho'\left(r,\theta\right)$ is the dynamical anomaly representing
fluid velocities with respect to the solid body rotation \citep{Kaspi2010a}. The zonal gravity harmonics that
represent only the dynamical part of the flow $\left(\Delta J_{n}\right)$
can be calculated by integrating the density anomaly and its projection
onto the Legendre polynomials in spherical coordinates such that
\begin{eqnarray}
\Delta J_{n} & = & -\frac{2\pi}{MR_{J}^{n}}\intop_{0}^{R_{J}}\intop_{-1}^{1}\rho^{\prime}\left(r,\mu\right)r^{n+2}P_{n}\left(\mu\right)d\mu dr.\label{eq:anom_Jn}
\end{eqnarray}

Since an oblate planet with no dynamics
is symmetric between north and south, the density anomaly represented
by the odd harmonics ($n=3,5,..$) should be identically zero if the
flow pattern is symmetric, and will be very small if the dynamics are shallow ($\Delta J_{n}=J_{n}$
for odd $n$). However, Juno measured four odd gravity harmonics \citep{Iess2018}, indicating the existence of a strong asymmetric pattern in Jupiter's flow field due to the existence of strong, deep winds.

The rapid rotation and size of the planet (small Rossby number) imply
that this asymmetry is directly related to zonal flows, since,
to first order, the leading balance in Jupiter is a geostrophic balance
between the flow-related Coriolis forces and the pressure gradients.
This leads to a vorticity balance known as thermal wind balance \citep{Pedlosky1987,Kaspi2009}.
If only zonal (azimuthal) flows are considered, the thermal wind balance
can be written as
\begin{eqnarray}
2\Omega r\frac{\partial\left(\tilde{\rho}u\right)}{\partial z} & = & g_{0}\frac{\partial\rho^{\prime}}{\partial\theta},\label{eq:thermal wind}
\end{eqnarray}
where $\Omega$ is Jupiter's rotation rate, $u\left(r,\theta\right)$
is the zonal flow,{ $g_{0}\left(r\right)$ is the mean gravitational
acceleration} and $z$ is the direction parallel to the rotation axis.
An equivalent equation can be written with temperature instead
of density gradients, and one can easily switch between the two versions {through the equation of state}. 
{Note that the barotropic limit of Eq. \ref{eq:thermal wind} is not simply when the rhs vanishes, but when $\frac{\partial u}{\partial z}=0$ (see full derivation at \citet{Kaspi2016}).} {\citet{Galanti2017a} showed that a higher order
expansion, beyond thermal wind, only slightly adjusts the deep flow dynamics 
(less than $10\%$). Therefore, for the purpose of studying the overall vertical profile, Eq. \ref{eq:thermal wind} is a good approximation.}

Our goal here is to search for possible deep wind structures that
can explain each of the measured odd gravity harmonics ($J_{3,\:}J_{5,\:}J_{7,}$
and $J_{9}$). Unlike previous studies \citep[e.g.,][]{Kaspi2018},
we are not solving for an optimal solution with respect to the full
error covariance matrix. Any vertical wind profile that fits the
odd measured gravity harmonics, within the uncertainty range of Juno,
is considered a possible solution for the flow. This allows us to examine
the full range of possible solutions, without converging on a single
decay profile of the flow. For example, the optimal solution suggested by
\citet{Kaspi2018} that considered the error covariance matrix is
not a solution here since the value of $J_{3}$ is not within the
measured error.

\section{The vertical profile of the zonal flow \label{sec:vertical_structure}}

Taking a hierarchal approach entailing an increasing level of complexity,
we first use the observed cloud-level wind as an upper boundary condition
for the flow field and assume the same profile continues {inward}
{in a direction parallel to the spin axis, due to angular momentum
considerations} \citep{Busse1976,Kaspi2010a}. The possible deep flow
structures are then set to decay continuously from the cloud level to a few
thousand kilometers below it \citep{Kaspi2018}, using two different decay
regions. Dividing the decay functions into two distinct
regions stems from the magnetic field's possible effects on the flow,
expected approximately at $r<0.97\,R_{J}$ \citep{duer2019,wicht2019a}, which imply
that once the electrical conductivity becomes dominant, the magnetic
field acts to dissipate the flow \citep{Liu2008,Gastine2014}. Thus,
for the lower part (the semiconducting region), we chose an exponential
decay {(Eq.~\ref{eq:simulation decay lower}, $r<R_{T}$)} that fits the exponential
nature of the electrical conductivity within Jupiter \citep{nellis1992,Weir1996,French2012}.
For the upper part, the vertical decay function includes both an exponent and hyperbolic
tangent {(Eq.~\ref{eq:simulation decay upper}, $R_{T}\leq r\leq R_{J}$)}, which combine to
give a wide range of possible decay profiles. 

The vertical profile of the zonal flow is defined with six independent
parameters, chosen to cover an extensive range of vertical profiles.
It is set as

\begin{equation}
u(\theta,r)=u_{{\rm proj}}(\theta,r)Q_{s}(r),\label{eq:simulation wind}
\end{equation}
\begin{equation}
Q_{s}(r)=(1-\alpha)\exp\left(\frac{r-R_{J}}{H_{1}}\right)+\alpha\left[\frac{{\rm tanh}\left(-\frac{R_{J}-H_{2}-r}{\Delta H}\right)+1}{{\rm tanh}\left(\frac{H_{2}}{\Delta H}\right)+1}\right]\qquad R_{T}\leq r\leq R_{J},\label{eq:simulation decay upper}
\end{equation}
\begin{equation}
Q_{s}(r)={Q_{s}(R_{T})}\exp\left(\frac{r-R_{T}}{H_{3}}\right)\qquad r<R_{T},\label{eq:simulation decay lower}
\end{equation}
{where $u_{{\rm proj}}(r,\theta)$ is the wind at the cloud level, projected inwards (with no decay) in the direction parallel to the axis of rotation
($\hat{z}$-axis}, Eq.~\ref{eq:thermal wind}), $Q_{s}(r)$ is the
radial decay function, representing the fraction of the cloud-level
wind at every depth, and the set of parameters that forms the decay
are bounded by the following limits: $0\leq\alpha\leq1$, $200$ km $\leq H_{1} \leq2500$
km, $200$ km $\leq H_{2} \leq2500$
km, $200$ km $\leq \Delta H \leq2500$
km, $0.95\leq R_{T}\leq0.975\,R_{J}$ and $100\leq H_{3}\leq900$
km.{ }The function $Q_{s}$ is also {smoothed} at the
transition depth. 
{We vary the parameters uniformly between their lower and upper bounds, taking only profiles where the wind speed decays monotonically as viable options. In total, we consider $5\times10^{5}$ decay profiles, sufficiently covering the parameter space. This set of decay profiles serves as the sample population for this study. For each profile, we calculate the associated density anomaly and the implied odd gravity harmonics.
All the calculations presented below are performed using the same sample population.} Note that other forms of $Q_{s}$ are possible, and can still fit the measured
gravity data ($Q_{s}(r,\theta)$, for example, as explored in \citet{Kaspi2018}). However, we have found that for the exploration
of the individual gravity harmonic depth sensitivities and the meridional
profile anomalies, the chosen function, which allows
a very wide range of decay profiles, is satisfactory.

\begin{figure}
\begin{centering}
\includegraphics[width=0.8\textwidth]{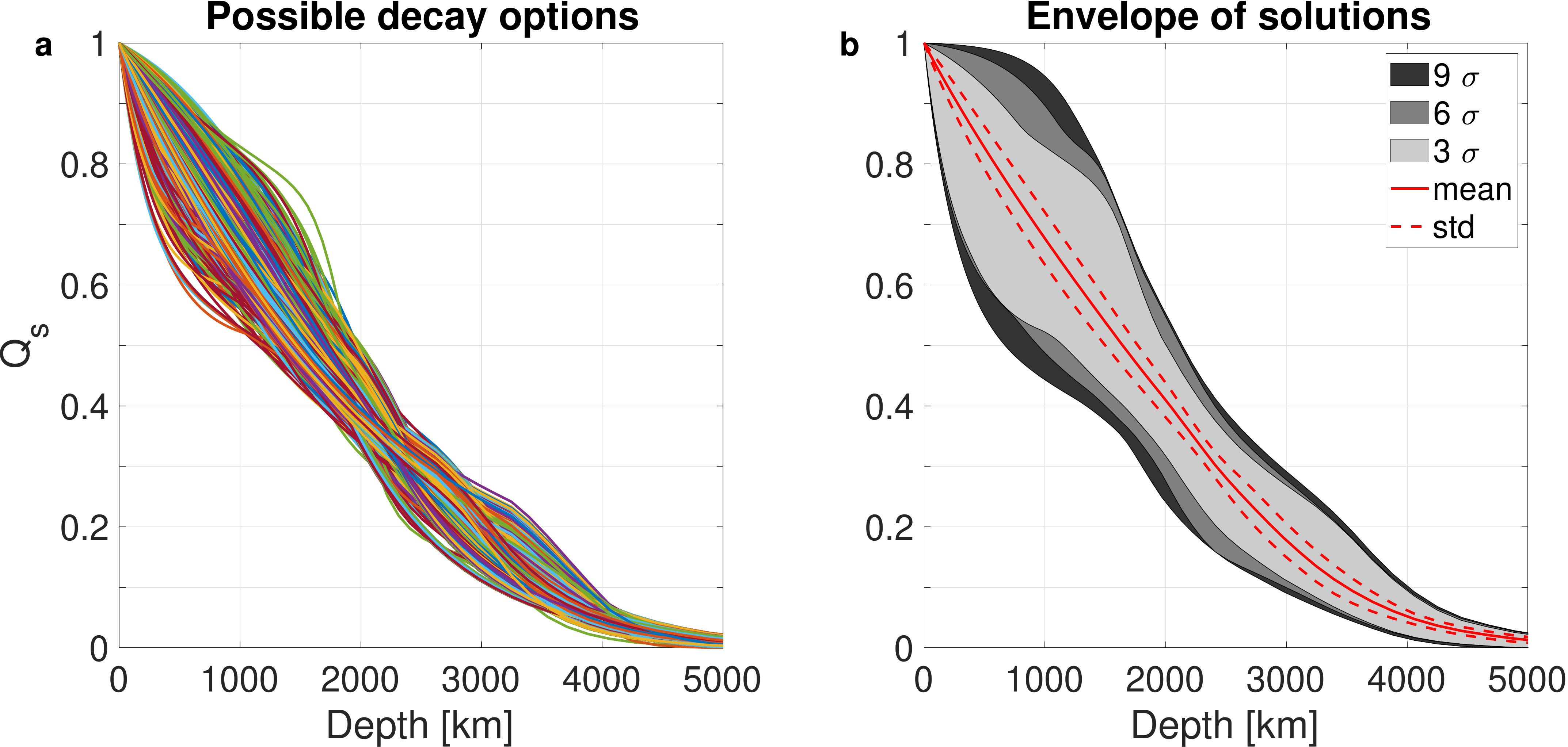}
\par\end{centering}
\caption{\label{fig:envelope of solutions} (a) Decay options that fit all
four measured odd gravity harmonics $\left(J_{n}\right)$ within the
$3\sigma$ sensitivity range of Juno, using Jupiter's observed cloud-level
flows \citep{Tollefson2017}. (b) Envelope of all possible solutions
(light gray), the average solution of all options (red line), their
standard deviation (dashed red lines) and envelope of solutions satisfying
larger {uncertainty} range (darker gray with rising uncertainty
range). }
\end{figure}

{From the }$5\times10^{5}${ decay options examined,
$6712$ vertical profiles are compatible with Juno's measured odd
gravity harmonics, which represent a little over $1\%$ of the sample
population (Fig.~\ref{fig:envelope of solutions}a). All the compatible
decay profiles are located in a relatively narrow envelope, especially in
the region around $2000$~km depth and  the one below $4000$~km, with all
the options pointing to remnants of jet-associated velocities
at a depth of $4000$~km (Fig.~\ref{fig:envelope of solutions}). Those deep
velocities are still on the order of $1$ ${\rm m~s^{-1}}$ and despite
being small, they are still higher than the magnetic secular variation
associated velocities estimated by \citet{moore2019}. Increasing
the error range of Juno's gravity measurements does allow for more
solutions, but the overall structure does not change much (Fig.~\ref{fig:envelope of solutions}b).}

\subsection{The depth sensitivity of the odd harmonics \label{subsec:depth_sens}}

\begin{figure}
\begin{centering}
\includegraphics[width=0.8\textwidth]{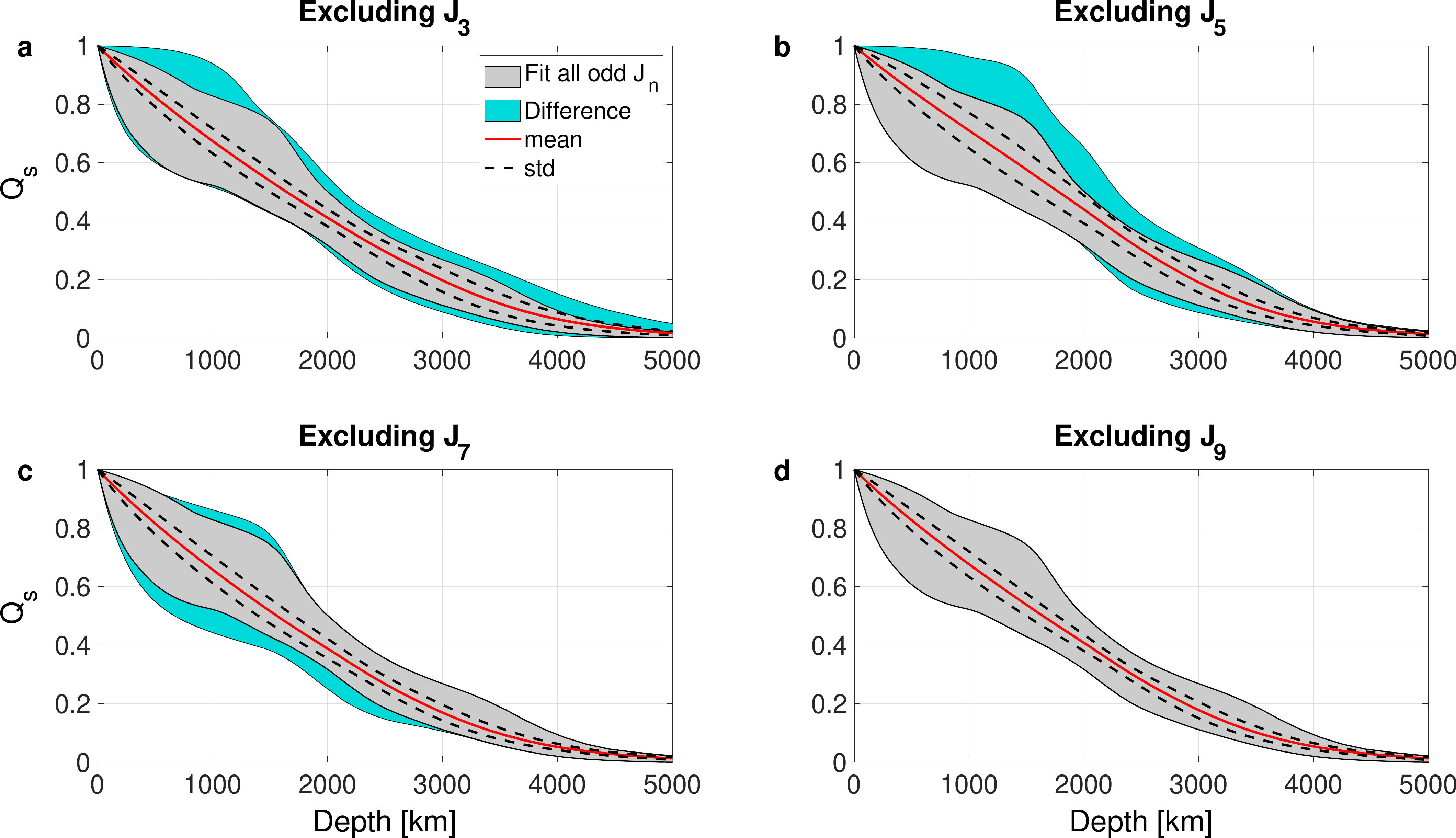}
\par\end{centering}
\caption{\label{fig:Jupiter wind envelopes} (a) The envelope of possible solutions
that fit all four odd gravity harmonics (gray), the envelope of additional
solutions once excluding $J_{3}$ (while still fitting $J_{5}$, $J_{7}$
and $J_{9}$) (turquoise), the average of all decay options within the
panel (gray and turquoise combined) (red), and their standard deviation (dashed black). The other panels are the same while excluding
other $J_{n}$: (b) excluding $J_{5}$, (c) excluding $J_{7}$ and
(d) excluding $J_{9}$. The results are a combination of the sample
decay options ($5\times10^{5}$). Note that the turquoise color emphasizes
the depth sensitivity of each gravity harmonic separately. }
\end{figure}

Research to date has focused on finding vertical profiles that match all
four odd gravity harmonics. However, there is information to be obtained
from each gravity harmonic separately. Here, vertical flow profiles
that fit three out of the four measured odd gravity harmonics are
considered, and the depth sensitivity of the excluded harmonic is
studied by examining the difference between the vertical profiles
that include the specific $J_{n}$ to those that {do not necessarily
include it.} The resulting depth sensitivity of each odd measured
gravity harmonic, according to Jupiter's measured zonal profile, is
presented in Fig.~\ref{fig:Jupiter wind envelopes}. The gray envelope,
the same one as in Fig.~\ref{fig:envelope of solutions}b, is
the boundary of all the possible solutions that fit all four odd gravity
harmonics within $3\sigma$. Note that not all possible profiles inside the gray
envelope will necessarily generate a solution compatible with the
measured gravity field{, since the solution is also dependent
on the decay profile within the given envelope.} All the other solutions
 gained while excluding one of the odd gravity harmonics appear
in Fig.~\ref{fig:Jupiter wind envelopes} (turquoise envelopes).
The turquoise envelopes always contain the gray envelopes by definition,
since they are constructed by fitting at least three gravity harmonics.
The difference between the turquoise envelopes and the gray ones denote
the region in which the \textit{excluded} harmonic bounds the flow.

The most insignificant influence is clearly of $J_{9}$ (Fig.~\ref{fig:Jupiter wind envelopes}d).
It appears to add no solutions at all to the gray envelope, meaning that
$J_{9}$ does not constrain the flow \textit{if} the other three odd
values are still within Juno's $3\sigma$. This is likely because
$J_{9}$ has the highest measurement error and lowest signal-to-noise
ratio (SNR), so even while fitting $J_{9}$, there is an extensive
region of solutions, and excluding it does not add new solutions.
The largest influence on the flow profile and depth sensitivity comes
from $J_{5}$ (Fig.~\ref{fig:Jupiter wind envelopes}b). It appears
to set the upper boundary of the gray envelope from the cloud level
($0$ km) to $3500$~km, and a lower boundary of the gray envelope
between $2000$ to $3500$~km. The strongest sensitivity is between
the cloud level and $3000$~km. $J_{5}$ has the smallest measured
$3\sigma$ value and {largest} SNR, but its value
is very similar to the SNR of $J_{7}$, so the large influence of
$J_{5}$ cannot be a result of the SNR alone. In a similar manner,
$J_{3}$ is mostly sensitive between $3000$ and $5000$~km and between
the cloud level ($0$ km) and $1500$~km (Fig.~\ref{fig:Jupiter wind envelopes}a).
Note that a flow profile that decays to zero at $4000$~km ($\sim0.94\,R_{J}$)
cannot fit $J_{3}$. $J_{7}$ is sensitive between $500$ and $2500$~km,
and sets mainly the lower boundary of the gray envelope at those depths
(Fig.~\ref{fig:Jupiter wind envelopes}c).

{Previous studies that examined the depth dependency of the even gravity harmonics (resulting from the shape and density of a solid-body model, without differential flows) concluded that higher-order harmonics are more sensitive to the density in the outer regions  \citep[e.g.,][]{zharkov1974,Guillot2007,Nettelmann2013b}. This is implied by the radial dependence of the gravity harmonics (Eq.~\ref{eq:Jn}). However, the above analysis shows that the wind-induced odd harmonics' depth dependency is more complicated. 
One exception is $J_3$, the only harmonic that is sensitive below $4000$ km, which resembles the even harmonics' depth tendency, where the low-order harmonics are more sensitive in deeper regions.}

\subsection{The contribution function \label{subsec:Contribution-function}}

The depth sensitivity of the gravity harmonics can also be examined
by calculating directly the depth dependence of $J_{n}${,}
{defined} as the contribution function{. This function}
was calculated in past studies for the even harmonics of Jupiter and
other planets \citep[e.g.,][]{Guillot2007,helled2010c,Nettelmann2013b}.
The contribution of each shell is the normalized integrant
of $J_{n}$, defined as 

\begin{equation}
C_{n}=\frac{1}{J_{n}}\frac{dJ_{n}}{dr}=\frac{1}{J_{n}}\frac{-2\pi}{MR_{J}^{n}}\intop_{-1}^{1}\rho\left(r,\mu\right)r^{n+2}P_{n}\left(\mu\right)d\mu\label{eq:contribution_eq}
\end{equation}
\citep{zharkov1974,hubbard1974,Hubbard1984}. The even harmonics in
{past} studies {were} calculated from the background
density (solid body models), while in our study we use the wind-induced
anomalous density field to calculate the odd harmonics' contribution,
taking $\rho^{\prime}$ instead of $\rho$ in Eq.~\ref{eq:contribution_eq}.

\begin{figure}
\begin{centering}
\includegraphics[width=0.8\textwidth]{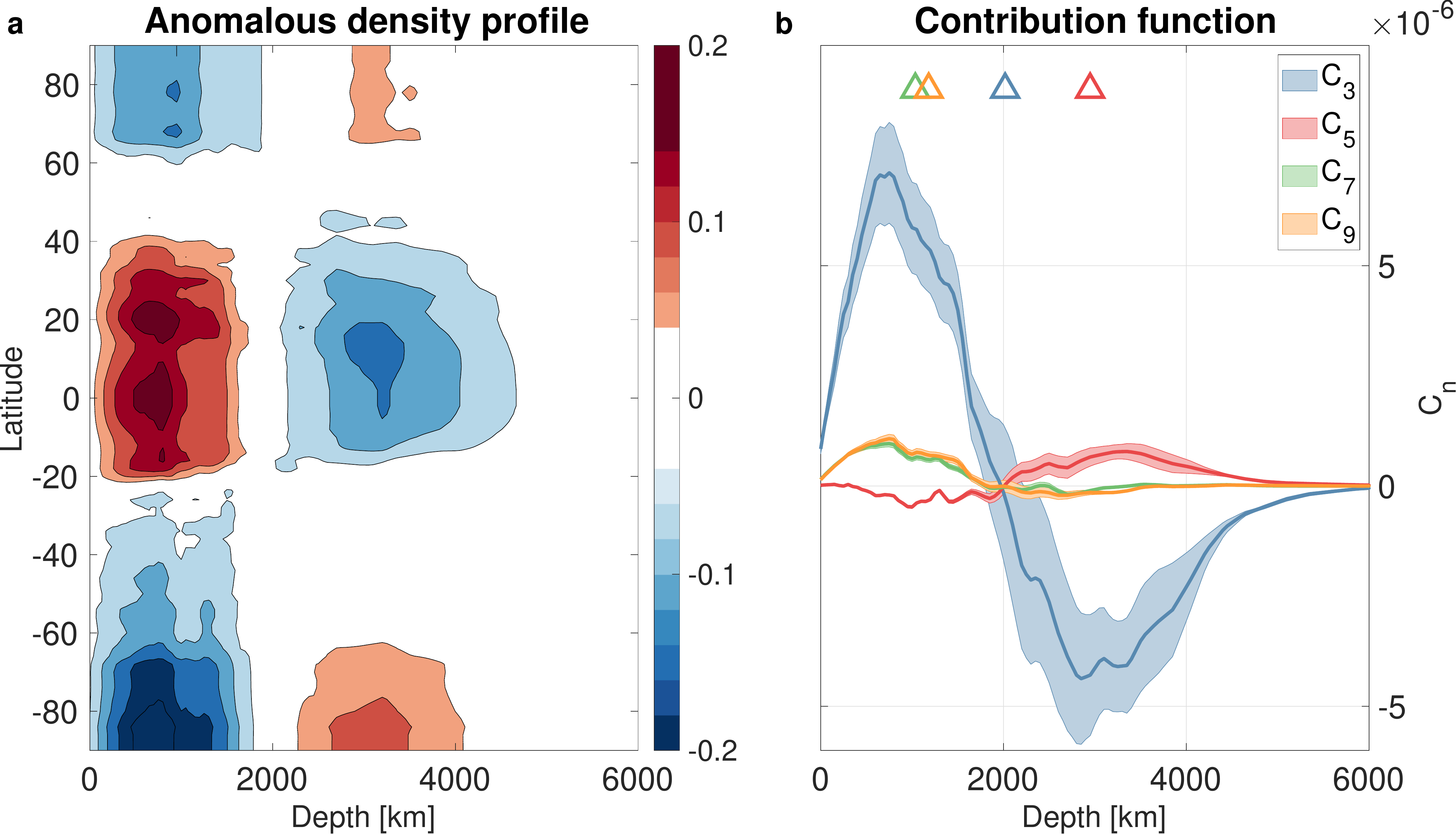}
\par\end{centering}
\caption{\label{fig:contribution} (a) {The mean anomalous density profile
of all possible decay options that fits the Juno four measured odd
gravity harmonics; colors represent the anomalous density values
$[{\rm kg\,~m^{-3}}]$. (b) Averaged contribution function (lines) $[{\rm m^{-1}}]$
for each of the odd gravity harmonics and their associated standard
deviation (shading); triangles represent the depth of the mean anomaly.
Both panels are for all the latitudes and for only the upper $\sim6000$~km,
below which the anomalous density is zero. }}
\end{figure}

{The averaged anomalous density profile of all possible decay
structures, that are consistent with the four odd gravity harmonics,
is presented in Fig.~\ref{fig:contribution}a. The anomalous density
reveals a change of sign at $2000$ km. The averaged odd contribution
functions ($C_{n}$) and standard deviations of each odd gravity harmonic
(Fig.~\ref{fig:contribution}b) corresponding to the solution envelope
from Fig.~\ref{fig:envelope of solutions} show a consistent sign
change.} {Note that the change of sign is exhibited only by the anomalous
density, corresponding to the wind shear with depth, and, therefore, does not exist when examining the even harmonics resulting from the static density} \citep[e.g.,][]{Nettelmann2013b}. {The integrals of the non-normalized contribution curve, $C_{n}$, are the gravity harmonic values, $J_{n}$, so the sign and value of $J_{n}$ are set by
the difference between the positive and negative curves (above and
below $2000$~km, not shown).} For the averaged anomalous density, the gravity
harmonics are: $J_{3}=-4.29\times10^{-8}$, $J_{5}=-7.50\times10^{-8}$, $J_{7}=10.8\times10^{-8}$ and $J_{9}=-6.69\times10^{-8}$.

The contribution function reveals a complex depth dependence for all
four gravity harmonics. {The depth sensitivity of each contribution function
is marked by the triangles (Fig.~\ref{fig:contribution}b), which
represent the depth of the mean absolute anomaly}{.}
{The contribution function of $J_{3}$, $C_{3}$, 
has the largest areas-under-the-curves at both
the shallower ($0-2000$ km) and deeper regions ($>2000$ km). The depth of the mean anomaly,
which here equals $2020$~km (Fig.~\ref{fig:contribution}b,
blue triangle), is near the depth of the sign change ($2000$~km),
meaning that $J_{3}$ gets near-equal anomalies from both regions.
{The standard deviation of $C_{3}$
(Fig.~\ref{fig:contribution}b, blue shading) is the largest, implying
a large variability of the solutions with depth when considering the $J_{3}$ value.} {The mean anomaly of $C_{5}$ is located in the deeper part of the domain
{(Fig.~\ref{fig:contribution}b, red triangle)}, and the standard deviation
of $C_{5}$ is substantial only between $2000$ and $4000$~km. $C_{5}$ is the only harmonic dominated mostly by the deeper region, emphasizing the important effect of $J_{5}$ on the deep wind structure (section \ref{subsec:depth_sens}).
The mean anomalies of $C_{7}$ and $C_{9}$ are clearly located in the shallower region {(yellow and green triangles $<2000$ km)}, {and their standard deviation}
{is small everywhere.} The contribution of both $C_{7}$ and $C_{9}$ is zero below $3000$ km, corresponding to Fig.~\ref{fig:Jupiter wind envelopes}. Since $C_{7}$ and $C_{9}$ lay almost on top of each other, $C_{7}$ might mask the depth dependency of $C_{9}$, as revealed in Fig.~\ref{fig:Jupiter wind envelopes} (along with the low SNR of $J_{9}$), so that if $J_{7}$ is within Juno's error range, so is $J_{9}$.
It is evident that the contribution function of the odd
harmonics exhibits a more complicated pattern than the classical even
harmonics \citep[e.g.,][]{Guillot2007,helled2010c,Nettelmann2013b}.
As in the previous analysis, we find that the higher-order odd harmonics are not simply
more pronounced in the outer regions. The projection of the wind patterns
onto different depths is reflected in the odd harmonics' contribution
at those depths, suppressing the $\left(r/R_{J}\right)^{n}$ dependency,
which is the prominent feature of the even harmonics' contribution.}}

\section{Sensitivity to the meridional profile of the zonal flows \label{sec:meridional}}

Next, we relax the assumption, used in the previous section, that the meridional profile of Jupiter's zonal flow
remains constant at all depths. First, the zonal wind profile is measured
by tracking cloud motion, which itself has some uncertainty \citep{Tollefson2017}. Second, and most importantly, the assumption that the cloud-level
profile extends perfectly to depth along the direction of the spin axis  requires the flow to be {locally close-to-barotropic} {(in the upper few thousand kilometers),
which is not necessarily the case.} {Although the flow cannot
be completely barotropic if $Q_{s}\neq1$ (Eq.~\ref{eq:simulation wind}),
the horizontal density gradients required to balance the vertical changes associated with $Q_{s}$
may be small (Eq.~\ref{eq:thermal wind}).} Any further
deviation from {close-to-barotropic flow} must be supported
by horizontal density (or temperature) gradients, which themselves must be
maintained by some internal mechanism \citep{Showman2013a}. Internal
convection models support the scenario that there may be internal
shear over the upper few thousand kilometers, but the overall structure
of the flow does not change much \citep{Kaspi2009,Jones2009}. Any
significant deviation from the zonal wind profile observed at the
cloud level requires significant shear and, therefore, notable horizontal
thermal gradients {(thermal-wind balance)}. As this is an open
question, for the purpose of this analysis, we examine several cases
of zonal wind meridional profiles, under the assumption that the wind
profile possibly varies {close to the cloud level and then
projects inward without further modifications}. For the purpose of
the gravity analysis, this means that the altered meridional profiles
occupy enough mass to affect the gravity field, and the flow observed at the cloud level
is limited to a shallow-enough layer so it does not
affect the gravity field.

The simplest case is clearly to use the measured profile at Jupiter's
cloud level and allow its magnitude to decay with depth (section \ref{sec:vertical_structure}).
A slightly less constraining option is to insert a perturbation into
the measured profile, thereby keeping the general form and allowing
a varying level of modifications to the cloud-level flow. The perturbed
winds chosen here might represent the measured uncertainties in Jupiter's
cloud-level wind \citep{garcia2001,Tollefson2017}. Finally, random
meridional profiles of the zonal flow with a spectra generally similar
to that of Jupiter are examined as well.

The modified zonal flow profile is chosen {at the cloud level
and projected inwards along the rotation axis ($u_{{\rm proj}}$,
Eq.~\ref{eq:simulation wind})} with a range of vertical profiles,
as described in section \ref{sec:vertical_structure}. The profiles
are calculated by adding sinusoidal perturbations to the measured wind. The modified profiles have a standard deviation
of $5\pm0.5\,{\rm m~s^{-1}}$ (varies with latitude) relative to the cloud-level winds, well within
the measurement error \citep{garcia2001,Tollefson2017}. The perturbation
is constructed as

\begin{equation}
\epsilon\left(\theta\right)={\sum}_{n=1}^{10}\left[a_{n}\sin\left(2n\theta\right)+b_{n}\cos\left(2n\theta\right)\right],\label{eq:perturb_eq}
\end{equation}
where $\epsilon$ is the perturbation and $a_{n}$ and $b_{n}$ are random
numbers that are normally distributed around zero with a {standard
deviation} of $2$ ${\rm m~s^{-1}}$. We first examine $1000$ modified profiles, each
constructed by adding the perturbation to the measured
wind (section \ref{subsec:Perturbed-zonal-profiles}). In addition, $1000$ random profiles are constructed purely
from the $\epsilon$ function (Eq.~\ref{eq:perturb_eq}), where $a_{n}$ and $b_{n}$ have a standard deviation of $30\,{\rm m~s^{-1}}$.
These profiles represent internal winds that are completely unrelated
to the observed cloud-level winds (section \ref{subsec:Random-zonal-profiles}).

\subsection{Perturbed cloud-level wind profiles \label{subsec:Perturbed-zonal-profiles}}

The perturbed wind profiles (Fig.~\ref{fig:semi jupiter depth sensitivity}a,
colors) result in a substantially bigger solution envelope (Fig.~\ref{fig:semi jupiter depth sensitivity}b.1-4,
gray) than the one from the measured zonal wind profile case, consistent
with the fact that a wider range of wind profiles is allowed. Note that
the overall shape has changed and that the flow can even vanish at
$\sim2500$~km. This might have an important implication, since the
initial time-dependent magnetic field results from Juno imply that
the wind at these depths are very weak \citep{duer2019,moore2019}.
An important result is that even for the perturbed winds there are no solutions that fit
at least three odd $J_{n}$ that vanish above $2000$~km. The
depth sensitivity of each harmonic is less pronounced than for the measured
wind case. This reflects the fact that Fig.~\ref{fig:semi jupiter depth sensitivity}
is a combination of all the possible solutions from $1000$ examined
meridional structures. Overall, $J_{3}$ is still 
{sensitive in the deeper regions (exemplified by the mean profile being weaker at depth, red line Fig. \ref{fig:semi jupiter depth sensitivity}b.1), although $J_{7}$ and $J_{9}$ contribute at depth as well. $J_{5}$ turns out
to be the most insignificant harmonic and $J_{9}$ does affect the
depth range of $1500-2000$~km, unlike in the unperturbed wind case.
The substantially larger range of solutions, however, does not manifest
in more solutions relative to the examined cases. From $1000$ examined
profiles individually tested with the decay sample population, only about
$0.1\%$ fit the anomalous gravity field compared to about $1\%$
in the unperturbed case (Fig.~\ref{fig:histogram plot}, red and
blue). {This suggests that although perturbed cloud-level wind profiles are possible, it is statistically more likely that a profile that is similar to
the projection of the observed cloud-level wind is indeed the profile in the
deeper atmosphere of Jupiter.}

\begin{figure}
\begin{centering}
\includegraphics[width=0.8\textwidth]{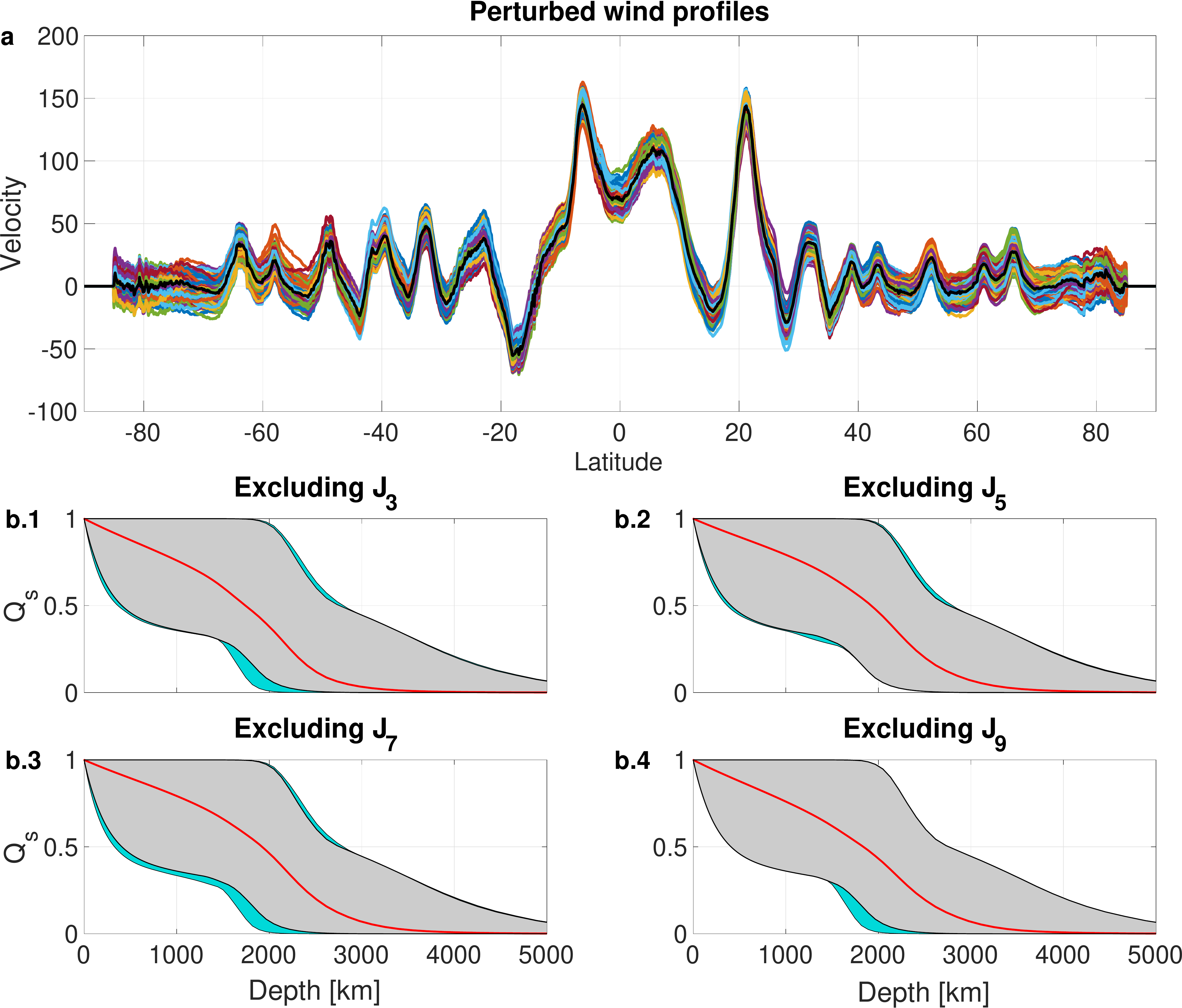}
\par\end{centering}
\caption{\label{fig:semi jupiter depth sensitivity} (a) 100 examples of the 1000 perturbed wind profiles (colors{, {[}${\rm m~s^{-1}}${]}}) and Jupiter's measured
wind profile (black). (b.1-4) The odd gravity harmonics of the perturbed wind profiles depth sensitivity summary as in Fig.~\ref{fig:Jupiter wind envelopes}. Each profile is examined with the same set
of decay options. The results shown here are for all meridional and
vertical options combined.}
\end{figure}

\subsection{The possibility of other zonal wind profiles \label{subsec:Random-zonal-profiles}}

Next, we consider profiles that do not resemble Jupiter's cloud-level
winds (Fig.~\ref{fig:Random pie charts}a). The resulting solution envelopes of the other zonal wind profiles
are relatively similar to the previous case of perturbed winds (not
shown). Only a very small subset of profiles
($13$ meridional profiles out of $1000$, about $1\%$) fit the four measured odd gravity harmonics (Fig.~\ref{fig:Random pie charts}b.1).
The possibility of fitting two or more odd harmonics is rare and exists in only $7\%$ or less of the zonal wind meridional profiles examined (Fig.~\ref{fig:Random pie charts}b.2-3).
$J_3$ is the harmonic that is pronounced in the majority of profiles 
(Fig.~\ref{fig:Random pie charts}b.4). In $14\%$ of the examined random profiles, no
odd harmonic is within the sensitivity range. In general, the measured harmonic's alignment with the zonal flow structure does not appear to be coincidental. {These findings are expected, considering that it is unlikely that an utterly different profile of zonal profiles arise below the cloud level of Jupiter.}

\begin{figure}
\begin{centering}
\includegraphics[width=0.8\textwidth]{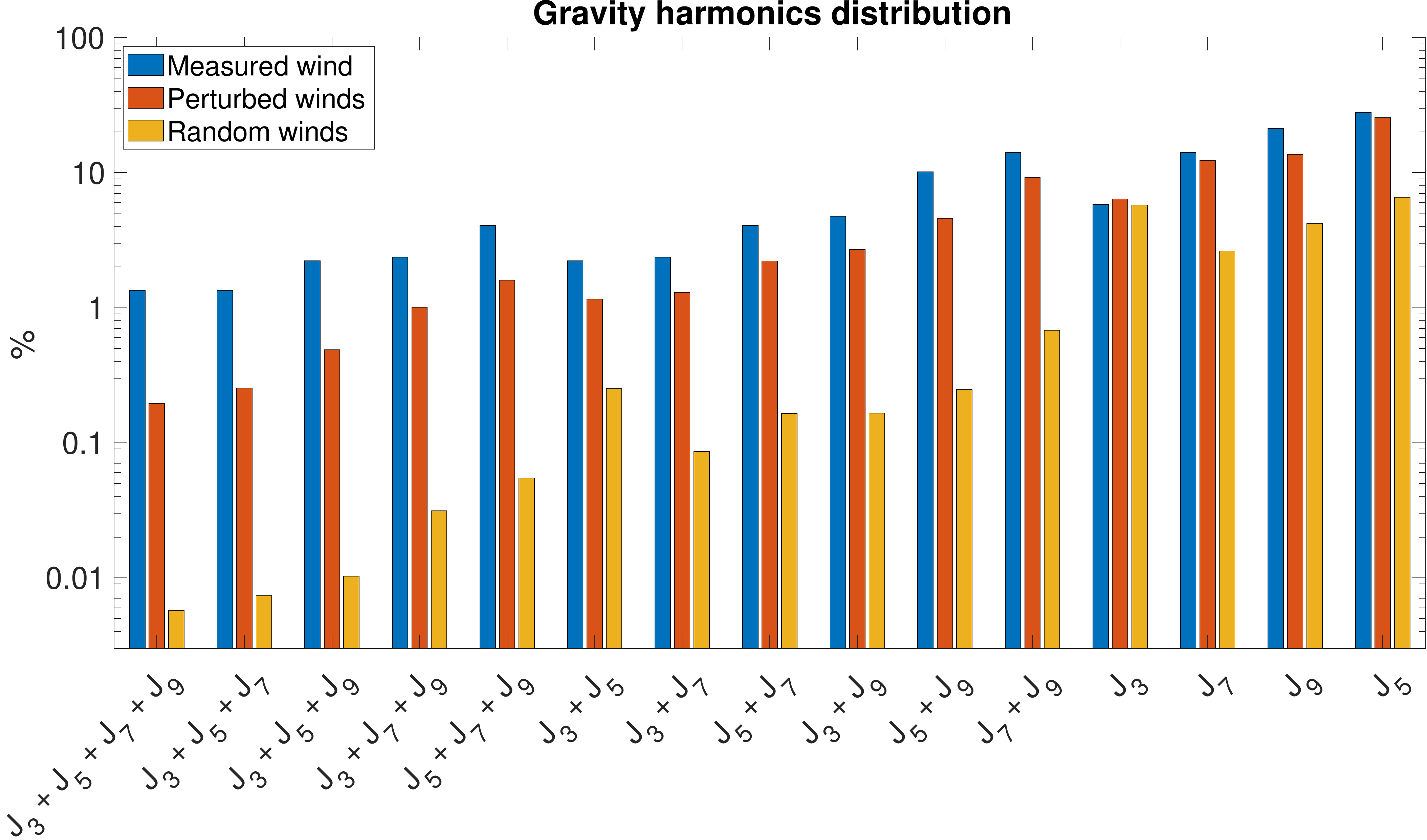}
\par\end{centering}
\caption{\label{fig:histogram plot} Summary of the solutions for the three presented
cases of wind structures: Jupiter's measured wind at the cloud level
(blue), 1000 slightly modified meridional structures (red) and 1000
random meridional profiles with a similar general structure to Jupiter's
meridional profile (orange). The ordinate is a logarithmic scale of percentage
relative to all the examined cases. The particular requirement
of the solution to match the different odd gravity harmonics is presented
by the abscissa.}
\end{figure}

{The ability of the $1000$ examined random profiles,
each with its sample of decay options, to fit all four odd gravity harmonics  is considerably smaller than
previous cases - only about $0.01\%$ (Fig.~\ref{fig:histogram plot}, orange). This indicates
that fitting all four odd harmonics is difficult with random meridional
profiles of zonal wind.} A summary of the examined cases appears in Fig.~\ref{fig:histogram plot}.
Note that the ordinate is a logarithmic scale and that $100\%$ stands
for all the zonal profiles ($1000$ zonal wind profiles other than
the measured cloud-level wind) and all decay options ($5\times10^{5}$) for each case. We find that the envelope of possible solutions from Fig.~\ref{fig:envelope of solutions}
stands for $\sim1\%$ of the tested vertical profiles for zonal
flows. The fitting percentage decreases with increasing perturbations,
and drops rapidly when switching to random profiles. This trend repeats
for all variations of {at least} three odd harmonics. For all
cases, the random winds show a significantly lower fitting percentage
than the other cases. We further present the fitting percentage obtained following the exclusion of
two and three harmonics.{ In summary, we find that
other meridional profiles of the zonal wind are possible, but
they are statistically unlikely. This result implies that the meridional profile of Jupiter's
zonal winds extends into the interior along the direction of the spin axis and weakens with
depth, and is likely not significantly different from the cloud-level profile. }

\begin{figure}
\begin{centering}
\includegraphics[width=0.8\textwidth]{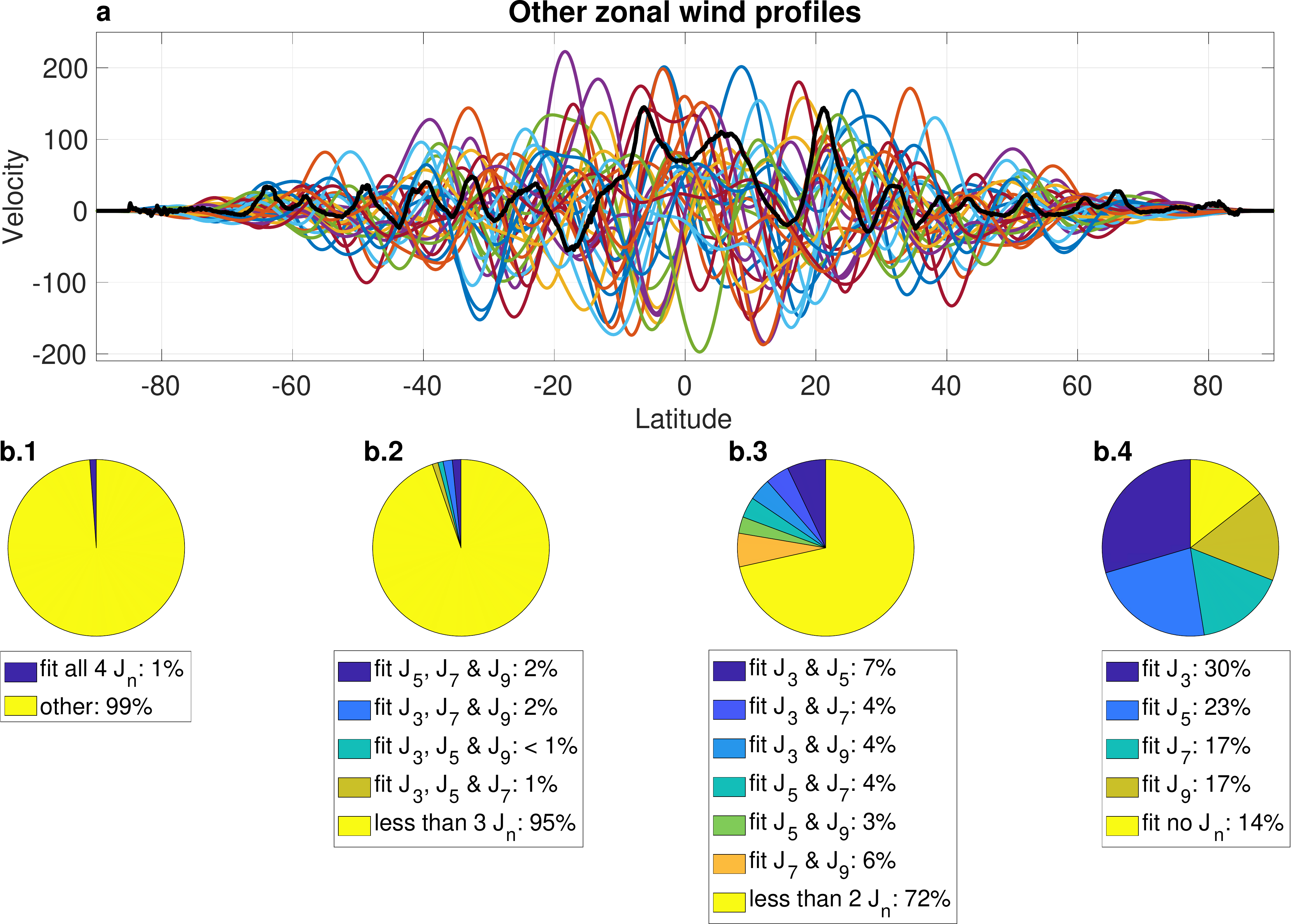}
\par\end{centering}
\caption{\label{fig:Random pie charts} (a) 30 examples of the 1000 random zonal wind profiles examined (colors{, {[}${\rm m~s^{-1}}${]}}) and Jupiter's measured
wind profile (black). (b.1-4) Summary of the random meridional profiles'
correspondence to the odd gravity harmonics. Only $\sim1\%$ of the
zonal profiles fit all four odd gravity harmonics (b.1),
$5\%$ of the zonal profiles fit at least three odd gravity harmonics
(b.2), $28\%$ of the zonal profiles fit at least two odd gravity harmonics
(b.3) and $14\%$ do not fit any of the odd gravity harmonics (b.4). The full
compatibility distribution is detailed in the figure.}
\end{figure}

\section{Zonal wind profiles inspired by the MWR measurements \label{sec:MWR}}

{In addition to the gravity measurements, 
Juno's six-channel microwave radiometer
(MWR) measurements might also reveal information about the structure of the wind below the cloud level. 
These measurements are used to calculate the nadir brightness temperature ($T_{b}$), a profile determined by the opacity of the atmosphere, which in Jupiter is determined mostly by ammonia abundance \citep{Li2017}. The MWR measurements reveal considerable
variation of $T_{b}$ with latitude and depth \citep{Bolton2017} (Fig. \ref{fig:MWR winds colors}a, black lines).
These variations with depth and the potential relation between $T_{b}$ and the zonal jets \citep{ingersoll2017b} suggest that Jupiter's zonal jets might be depth-dependent, similarly to $T_{b}$, instead of simply projected inwards (as in sections \ref{sec:vertical_structure} and \ref{sec:meridional}).}

{One approach for describing the relation between the brightness temperature and the zonal jets is taking the
brightness temperature as simply temperature. Then, the relation is described by the thermal wind
balance, as discussed in section \ref{sec:Methodology}. This approach, however, results in equatorial wind that is greater by two orders of magnitude than
the measured cloud-level wind, which is unrealistic \citep{Bolton2017}.
Another approach is taking the brightness temperature as an indicator for ammonia concentration \citep{ingersoll2017b} and examining the relation to the zonal jets. {As an example, such a relation is expected in the meridional circulation (Ferrel cells), where the cell-associated vertical velocity redistribute the substance and is accompanied by zonal jets \citep{fletcher2020}}. Here, we take the latter approach, analyzing a range of depth-dependent meridional profiles, compatible with the brightness temperature variations with depth.}

{When examining the correlation between $T_{b}$ and the zonal jets, a different analysis should be taken at different latitudes, and perhaps at different depths.
If the zonal jets are associated with multiple Ferrel cells in alternating directions associated with regions of momentum convergence (eastward jets) and divergence (westward jets), a correlation is expected between the zonal velocity and the ammonia concentration gradient. In such a scenario, the meridional cells advect the ammonia concentration, maximizing its gradient where the jet peaks \citep{fletcher2020}. However, momentum fluxes converging at the equator would lead to a superrotating jet \citep{Kaspi2009} and might also lead to a maximal ammonia concentration. Therefore, at the equator, the zonal velocity is associated with the concentration itself and not with its gradient. These simple considerations motivates us to examine the correlation both between $u$ and $\nabla T_b$ and between $u$ and $T_b$ (Table \ref{tab:Correlation-coefficients-between}, columns 2 and 3). Note that, as in all our experiments, the zonal jets are projected inwards along the spin axis, as in section \ref{sec:vertical_structure} (Fig. \ref{fig:MWR winds colors}a, colors). It is evident that the correlation between $u$ and $T_b$ is weak at the cloud level (channel 6, $0.6$ bar), but become stronger with depth (maximum at channel 1, $240$ bar), while the correlation between $u$ and $\nabla T_b$ is strong at the cloud level, and weakens with depth (at channels 1-3, the correlation is weak). This alone might indicate two opposite meridional cells, one stacked on top of the other \citep{Showman2005, fletcher2020}. At the cloud level, the correlation between $u$ and $\nabla T_b$ improves if we do not consider the equatorial region, which is consistent with the Ferrel cells hypothesis. Projecting the winds in the radial direction instead of along cylinders does not improve the correlation to neither $T_b$ or $\nabla T_b$ (Fig.~\ref{fig:MWR winds colors}b.1).}

{The dominant feature leading to the strong correlation between $u$ and $T_b$ at channel 1 is the equatorial anomaly, ascending at $\sim 15^\circ~\rm{S}$ and descending at $\sim 15^\circ~\rm{N}$ (Fig.~\ref{fig:MWR winds colors}b.1). While at the cloud level, both the zonal jets and $T_b$ reveal
alternating patterns (Fig.~\ref{fig:MWR winds colors}b.6), the waviness of $T_b$ vanishes at deeper depths (Fig.~\ref{fig:MWR winds colors}b.1-2). Since $T_b$ is depth-dependent, getting smoother with depth from
channel 6 (cloud level) to channel 1 ($\sim300$~km depth), we examine zonal jets that are depth-dependent. {Note that winds, projected along the spin axis, maintain their meridional profile with cylinders, and without further assumptions, are not depth-dependent}.
The modified (smoothed) wind at channel 1 is composed using a running average
of $\Delta\theta$ degrees latitude, where $\Delta\theta=0,1,2,...,10^\circ$ ($0^\circ$ means that no running average is applied). The wind at channel 6 is the observed cloud-level profile; between channel 1 and channel 6, the wind strength is interpolated. Finally, the wind profile at the depth of $300$~km (channel 1) is projected inwards along the spin axis with a decay profile as in the previous sections without further assumptions.} 
In addition to the projected winds with no depth-dependency ($\Delta\theta=0^{\circ}$, Fig.~\ref{fig:MWR winds colors}b, light blue), we examine the correlation between $u$ and $T_b$  for two cases of a depth-dependent zonal wind ($\Delta\theta=4^{\circ}$ and $\Delta\theta=8^{\circ}$, Fig.~\ref{fig:MWR winds colors}b, blue). Increasing the running average at depth improves the correlation at channels 1-3 (columns 3-5, Table \ref{tab:Correlation-coefficients-between}), implying that the latitudinal variability of the jets might weaken beneath the cloud level.}

\begin{figure}
\begin{centering}
\includegraphics[width=1\textwidth]{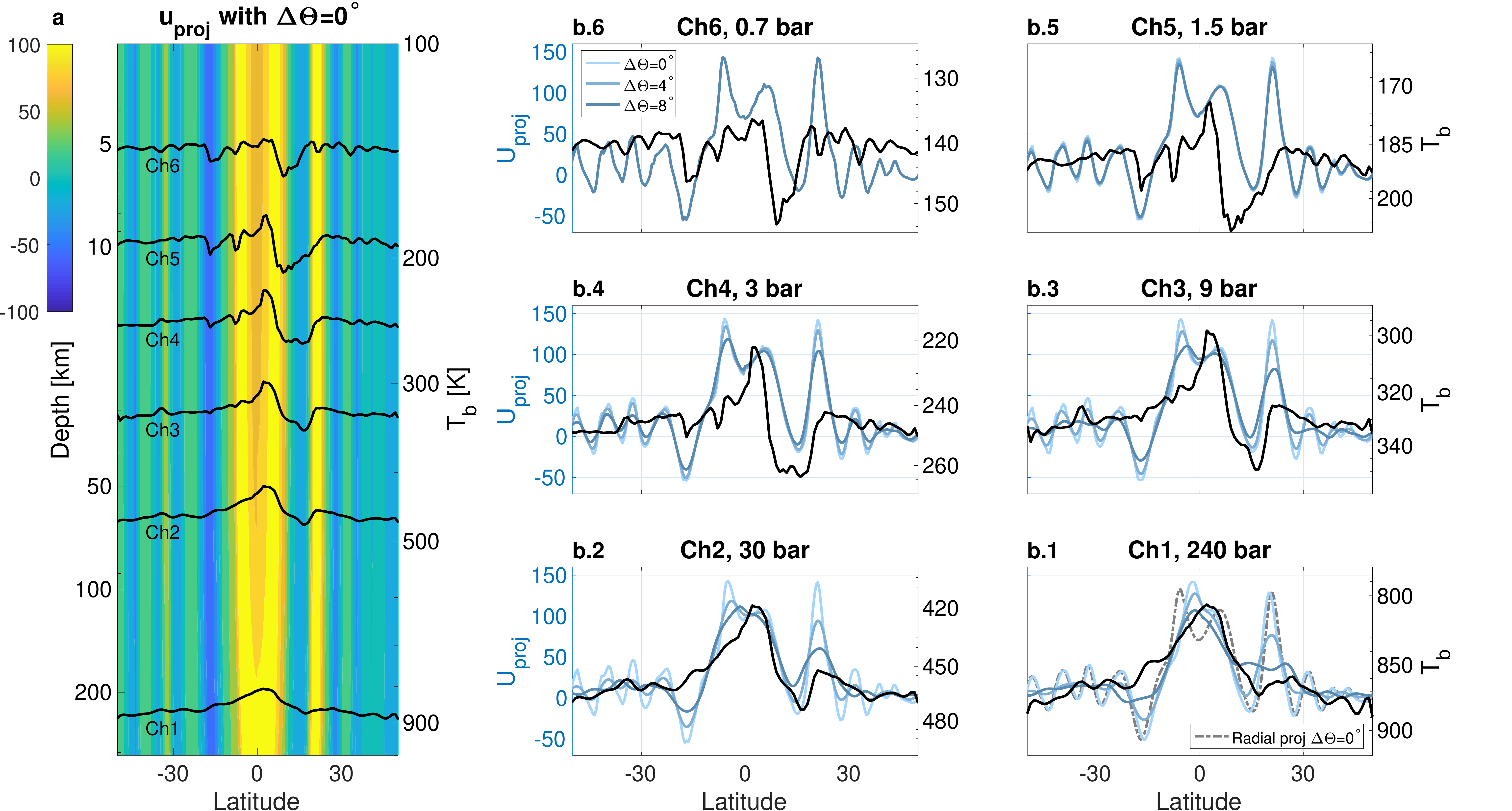}
\par\end{centering}
\caption{\label{fig:MWR winds colors} {(a) Jupiter's projected wind
velocities (colors, {[}${\rm m~s^{-1}}${]}) between latitudes $-50^{\circ}$
and $50^{\circ}$ in the upper $300$~km of Jupiter (left ordinate)
combined with nadir brightness temperature lines from Juno's PJ1 (black,
right ordinate) in channels 1 to 6, associated with frequencies of
$0.6$, $1.2$, $2.6$, $5.2$, $10$ and $22$ ${\rm GHz}$, respectively. (b.1-6) 
Jupiter's projected wind velocities [${\rm m~s^{-1}}$] at  channel
1 , $240$~bar (b.1), channel 2 , $30$~bar (b.2), channel 3, $9$~bar (b.3),
 channel 4, $3$~bar (b.4), channel 5, $1.5$~bar (b.5), and 
channel 6, $0.6$~bar (b.6) for a running average of $\Delta\theta=0^{\circ}$,
$\Delta\theta=4^{\circ}$ and $\Delta\theta=8^{\circ}$ (darker blue
with increasing $\Delta\theta$, left ordinate). Also shown is the brightness
temperature [$^{\circ}{\rm K}$] (black, right ordinate). The radial
projection of the winds with no running average (dashed gray) is also
presented in b.1.}}
\end{figure}

\begin{table}
\caption{\label{tab:Correlation-coefficients-between} {Correlation coefficients
 between $\nabla T_b$ and the wind velocity at each channel, for winds with no running average at depth
($\Delta\theta=0^{\circ}$), and correlation coefficients
 between $T_b$ and the wind velocity at the same depths
with no running average ($\Delta\theta=0^{\circ}$),
with running average of $4$ degrees ($\Delta\theta=4^{\circ}$),
and with a running average of $8$ degrees ($\Delta\theta=8^{\circ}$).
Note that for $u$ vs. $T_{b}$, the correlation increases with depth (or decreases with channel)
and with running average.}}
\centering
\begin{tabular}{c | c | c c c} 
\hline
      &       {$u$ vs. $\nabla T_{b}$} & \multicolumn{3}{c}{$u$ vs. $T_{b}$} \\
\hline
Channel & $\Delta\theta=0^{\circ}$ &  $\Delta\theta=0^{\circ}$ & $\Delta\theta=4^{\circ}$ & $\Delta\theta=8^{\circ}$\\
\hline
  $1$   & $0.05$ & $0.64$ & $0.76$ & $0.86$  \\
  $2$  & $0.02$ & $0.74$ & $0.80$ & $0.84$     \\
  $3$  & $0.01$ & $0.63$ & $0.65$ & $0.68$     \\
  $4$  & $0.15$ & $0.41$ & $0.41$ & $0.42$     \\
  $5$  & $0.24$ & $0.13$ & $0.13$ & $0.13$     \\
  $6$  & $0.42$ & $0.10$ & $0.10$ & $0.10$     \\
\hline
\end{tabular}
\end{table}

\begin{figure}
\begin{centering}
\includegraphics[width=0.8\textwidth]{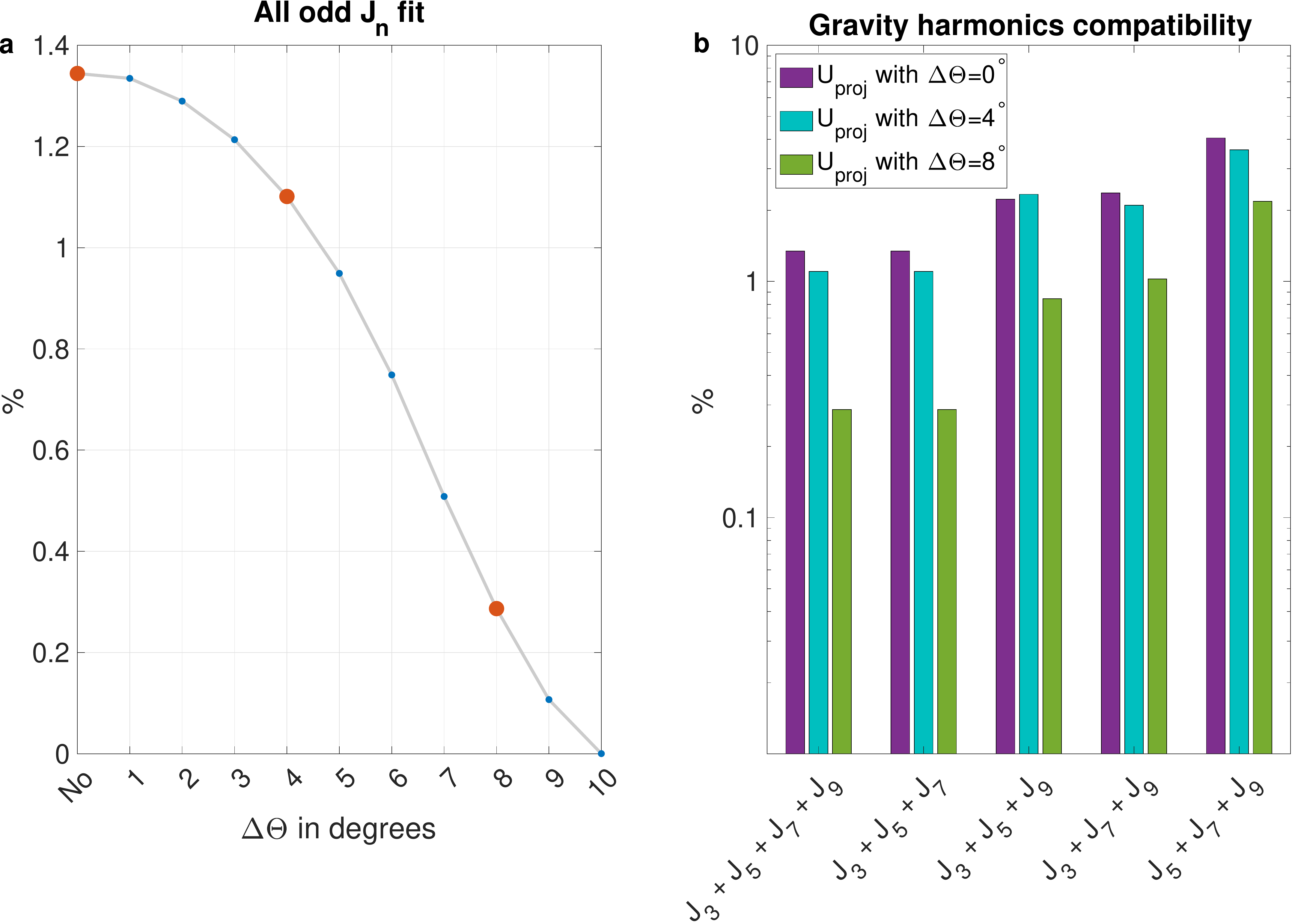}
\par\end{centering}
\caption{\label{fig:MWR results} (a) The ability of the depth-dependent wind
profiles to fit all four odd gravity harmonics (percentage of solutions)
as a function of the smoothing factor in degrees (blue line). The three
cases shown in Fig.~\ref{fig:MWR winds colors}b are denoted by red
dots. (b) The gravity harmonics distribution for the red dots is compatible
with the three case studies in Fig.~\ref{fig:MWR winds colors}b. The
ordinate is a scale of percentage relative to the $5\times10^{5}$ decay options examined.}
\end{figure}

{Next, we examine the ability of the depth-dependent zonal profiles to explain the measured odd gravity harmonics. We examine a range of case studies, from slightly
to largely modified depth-dependent profiles, until no solutions are
found (Fig.~\ref{fig:MWR results}a). For slightly smoother profiles (small $\Delta\theta$),
the ability to fit all four odd $J_{n}$ is similar to that without
any smoothing ($\Delta\theta=0^{\circ}$) (Fig.~\ref{fig:MWR results}a). Applying additional smoothing (increasing $\Delta\theta$)
decreases the ability to fit the four odd $J_{n}$. Using more than a
$10$-degree running average results in no solutions for the odd
gravity harmonics. The three case studies ($\Delta\theta=0^{\circ}$, $\Delta\theta=4^{\circ}$ and $\Delta\theta=8^{\circ}$) show a consistent trend when excluding one of the odd harmonics, such that the ability to fit the gravity measurements is reduced when it comes to smoother deep profiles (Fig.~\ref{fig:MWR results}b).
{This result is compatible
with the previous case (section \ref{sec:meridional}), indicating that deep wind that resembles
the cloud-level wind can fit the gravity data, and changing the
zonal wind structure considerably limits the ability to find a solution.}
The main conclusion of the analysis presented above is that wind profiles correlated with $T_b$ at depth ($\Delta\theta=4^{\circ}$) can adequately fit the gravity measurements.}

\section{Discussion and conclusions\label{sec:Conclusions}}

The main challenge of interpreting the Juno gravity measurements is
that the measurements provide only a handful of numbers (gravity harmonics),
while the meridional and vertical profile of the interior flow have
many degrees of freedom. Therefore, by-definition, the problem is
ill-posed. Acknowledging this inherent issue, \citet{Kaspi2018} used
four degrees of freedom for the vertical flow profile (matching the
number of the four odd harmonics), and found the best optimized profile
for this allowed range. They addressed the non-uniqueness by showing
the statistical likelihood of wind profiles for the interior that
are completely different than the cloud-level flow. \citet{Kong2018}
highlighted the non-uniqueness issue by showing that two different
flow profiles can still satisfy the gravity measurements. In this
study, we take a more methodological approach and consider a wider
range of solutions and analyze their statistical likelihood. The flow
profiles we consider, both for the meridional and vertical profiles,
are bound by physical considerations. We also address two main issues:
First, all previous studies looked at all four odd gravity harmonics
together, and found the flow profiles best matching all four. Here,
we investigate how each one of them separately bounds the flow.
Second, in an attempt to coincide the gravity and microwave data, we
explore whether deep profiles that are smoother than those of the cloud-level, 
as possibility indicated by the Juno microwave measurements, can be consistent with the gravity measurements.

{By assuming that the cloud-level wind profile
is projected inwards parallel to the spin axis}, with some decay profile{,}
we identify the envelope of possible solutions (Fig.~\ref{fig:envelope of solutions}).
We then relax the dependence on each of the odd gravity harmonics
separately and analyze their individual contribution to the
vertical profile of the zonal wind (Fig.~\ref{fig:Jupiter wind envelopes}).
We find that $J_{3}$, the lowest order odd harmonic that
represents the dynamics of Jupiter, is sensitive at depths where the
conductivity rises (beyond $\sim3000$~km), and the magnetic field
might be interacting with the flow, resulting in the Lorentz force
playing a key role in the dynamics. $J_{5}$ appears to be the most
sensitive harmonic, giving a robust constraint on the vertical profile
of the zonal flow alone (Fig.~\ref{fig:Jupiter wind envelopes}b).
Interestingly, $J_{9}$ does not give any new constraint on the flow
if the other three harmonics are within the sensitivity range (Fig.~\ref{fig:Jupiter wind envelopes}d).
{A possible explanation for the unique nature of $J_{5}$ comes from
exploring the contribution function, which revealed that $J_{5}$ is most
sensitive in the deeper regions, below $2000$~km }(Fig.~\ref{fig:contribution}){.}

The modified zonal flow analysis revealed a substantially bigger
possible solution envelope than that obtained by extending the cloud-level
wind (Fig.~\ref{fig:semi jupiter depth sensitivity}). This implies
that the zonal wind's structure may influence the depth sensitivity of each harmonic.
Even for the perturbed winds, the flow cannot vanish at depths shallower than
$2000$~km. The case with random winds implies that, {with
high probability}, the wind cannot alter completely below
the cloud level. Fitting the four odd gravity harmonics (or
three if we ignore $J_{9}$) requires either
similar winds to the measured ones  at the cloud level, that would penetrate a few thousand kilometers into the planet, or a very specific and statistically
unlikely combination of a meridional and a decay profiles
(Fig.~\ref{fig:histogram plot}, \ref{fig:Random pie charts}). Finally, the gravity harmonics
induced by the slightly modified depth-dependent meridional profiles, which
have a better correlation with the MWR measurements at depth (Fig.~\ref{fig:MWR winds colors}, Table \ref{tab:Correlation-coefficients-between}),
are still within Juno's gravity measurements' uncertainty, indicating that
Jupiter's ammonia abundance could indeed reflect the profile
of the zonal jet at 300~km (Fig.~\ref{fig:MWR results}).

\textit{Acknowledgments:}

We thank Cheng Li for providing the MWR data. This research has been
supported by the Israeli Space Agency and the Helen Kimmel Center
for Planetary Science at the Weizmann Institute of Science. The Juno gravity measurements and MWR measurements are publicly
available, see \break https://pds-atmospheres.nmsu.edu/data\_and\_services/atmospheres\_data/JUNO/juno.html.
Additional data can be found here https://doi.org/10.5281/zenodo.3859828.

\bibliographystyle{apalike}




\end{document}